\documentstyle[prb,aps]{revtex}
\input epsfig.sty

\begin{document}
\advance\textheight by 0.2in
\twocolumn[\hsize\textwidth\columnwidth\hsize\csname@twocolumnfalse%
\endcsname

\begin{flushright}
{\tt to appear in Phys. Rev. B } 
\end{flushright} 

\title{Dynamical transitions and sliding friction in the two-dimensional
Frenkel-Kontorova model}
\author{Enzo Granato}
\address{Laborat\'{o}rio Associado de Sensores e Materiais, \\
Instituto Nacional de Pesquisas Espaciais, \\
12201 - S\~{a}o Jos\'{e} dos Campos, SP, Brazil}
\author{S.C. Ying}
\address{Department of Physics, Brown University,\\
Providence, RI 02912, USA}
\maketitle

\begin{abstract}
The nonlinear response of an adsorbed layer on a periodic substrate
to an external force is studied via a two dimensional uniaxial 
Frenkel-Kontorova model. The nonequlibrium properties of the model 
are simulated by Brownian molecular dynamics. Dynamical phase 
transitions between pinned solid,
sliding commensurate and incommensurate solids  and hysteresis effects are
found that are qualitatively similar to the results for a Lennard-Jones
model, thus demonstrating the universal nature of these features.
\end{abstract}

\pacs{68.35.Rh, 68.35.Gy}

]

\section{Introduction}
In recent years, there have been an increasing effort
at trying to understand the frictional force  between
two macroscopic sliding surfaces at the microscopic level
\cite{Israelachvili,Persson,conf,book,granato,braiman,braun,kawaguchi,weiss}.
It is now understood that in the case of boundary lubrication, the tightly
bound monolayer in the regions of closest contact between the two surfaces
plays a crucial role in determining the macroscopic sliding friction. At
the same time, new experimental techniques such as the quartz-crystal
microbalance (QCM)\cite{Krim} and atomic force microscope (AFM)
\cite{conf} allow the direct probing of frictional force at the atomic
level. Thus, the central theoretical issue in understanding boundary
lubrication and atomic scale friction is to determine how an adsorbed
layer responds to an external driving force. In the linear regime, the
response  is characterized by the collective diffusion constant which is
itself an important transport coefficient in determining surface dynamics
such as growth of thin films \cite{growth}. The nonlinear regime is the
one relevant for actual conditions of macroscopic sliding motion. Persson
\cite{Persson} has made a series of pioneering studies on this problem
based on a model of an adlayer of particles interacting with Lennard-Jones
potentials. The central result is that there are intriguing dynamic phase
transitions between pinned solid, sliding solid and liquid phases.
Strong hysteresis effects exist in the transitions between these phases,
leading  to the 'stick and slip' motion often observed in the sliding
motion between two macroscopic flat surfaces under  an external driving
spring. In addition, Persson has used  a general hydrodynamic argument
to predict that the ratio of the static and sliding friction threshold
should take a universal value of approximately 2 for all systems 
with strong coupling to the substrate.

The qualitative features of the results from this isotropic Lennard-Jones
system are very encouraging. Besides the relevance for the boundary
lubrication problem, the dynamic phase transitions that have been observed
are of intrinsic interest and very different from the  more familiar
equilibrium transitions. However, actual adsorbate layers usually have
strong anisotropies and complicated many-body interactions among the
adatoms. The question then arises as to how universal are the results
for the Lennard-Jones system.  Recently, Braun {\it et al} \cite{braun} have
studied the nonlinear mobility of an adlayer based on a generalized vector
Frenkel-Kontorova model. They have also observed a strong hysteresis
effect in the transition between the pinned and running state,
while the nature of the running state and the ratio of the static to
sliding friction threshold they obtained are quite different from the
corresponding results of Persson.  However, it is difficult to make a
direct comparison between these studies. Persson focused mainly on an
initial pinned state that is commensurate with the substrate, while Braun
{\it et al} focus almost exclusively on an initial state that is incommensurate
with the substrate.  In the latter case, the sliding motion is dominated
by the response of the kinks or antikinks to the external force.

To facilitate a more direct comparison with Persson's study and address
the question of universality in the salient features, we present in
this paper the results from a study based on a uniaxial Frenkel-Kontorova 
model that is different from the one employed by Braun {\it et al}. 
In particular, the particles in our model have  harmonic spring-like 
interactions that prevents them from moving away from each other. 
This implies a fixed
coordination number for each adparticle and a very different nature of
the liquid phase from the corresponding phase for the Lennard-Jones
system. Moreover, the particle displacement is a scalar variable in 
contrast to the vector displacements in the other models. Also, in 
this paper we focus mainly  on an initial pinned state that is 
commensurate with the substrate. Surprisingly, we find that almost 
all the features of the dynamic phase transitions and hysteresis 
effects found in the Lennard-Jones system are present also for our 
generalized Frenkel-Kontorova model. 
The only notable difference between the models is in the value of 
the ratio of the static to sliding friction threshold. We present 
below the details of the model and the results. Preliminary results 
have been reported in an earlier conference proceeding \cite{granato}.

\section{Model}

We consider a two-dimensional version of the Frenkel-Kontorova (FK)
model, which is the simplest extension of the one-dimensional model
\cite{Frenkel}, obtained by coupling linear harmonic chains along the
additional dimension. The particle displacement is always along the
same direction which makes it particularly convenient for numerical
computations. This model is clearly more appropriate to describe uniaxial
systems of adsorbed layers \cite{Pokrovsky} however we believe that the
main features of the dynamical behavior which we consider in this work
should be independent of this type of anisotropy. We have in fact checked
that the dynamical transitions and hysteresis effects remains essentially
the same if a more isotropic version of the model \cite{shinjo} is used.

\begin{figure}[tbp]
\centering\epsfig{file=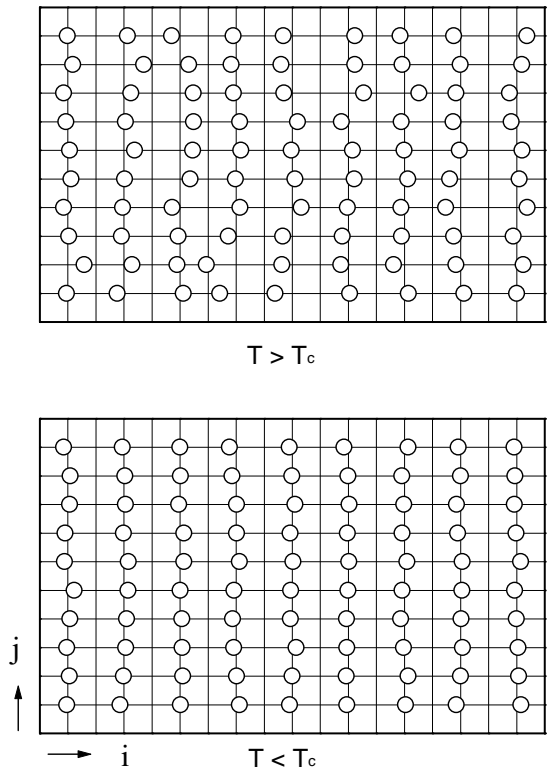,bbllx=6cm,bblly=3cm,bburx=16cm,
 bbury=26cm,width=4.5cm}

\caption{Snapshot pictures of the adsorbate above 
and below the depinning temperature $T_c$. (a) $T=3$ and (b) $T=0.5$. }
\label{config}
\end{figure}

The model is described by the Hamiltonian

\begin{eqnarray}
H&=&\sum_{i,j}\ [ \frac{p_{ij}^{2}}{2m}+\frac{K_{x}}{2}
(x_{i+1,j}-x_{ij}-b)^{2} \nonumber \\
&&+\frac{K_{y}}{2}(x_{i,j+1}-x_{ij})^{2}-U_{o}\cos (
\frac{2\pi x_{ij}}{a})]  \label{hamilt}
\end{eqnarray}
where $m$ is the mass of the particles; $U_{o}$ the amplitude of the
periodic potential with period $a$ ; $K_{x}$ and $K_{y}$ are elastic
constants in the x and y direction and $b$ is the average distance between
the particles in the absence of the external potential. The second term
in the brackets represents the elastic interaction within a chain and
the third term the interaction between adatoms on different chains. Note
that the index $i,j$ labels the particles within a chain with integer $i$
($i=1,...,$ $N_{x}$ ) and the chain with integer $j$ ($j=1,...,$ $N_{y}$
), so that the system consists of a total number of $N_{x}\times $
$N_{y}$ particles. In the simulations we have considered $N_{x}=$
$N_{y}=N$ and $K_{x}=K_{y}=K$. Typical equilibrium particle configurations 
above and below the depinning temperature $T_c$ are illustrated 
in Fig. \ref{config} for $b/a=2$ 

We use periodic boundary conditions in most of the calculations. This
is obtained by requiring that the particle $(N_{x}+1,$ $j$ $)$ is
a periodic image of the particle $(1,j)$ in the same chain and so $
x_{N_{x}+1,j}=x_{1,j}+N_{s}\ a$, where $N_{s}$ is the number of local
minima of the periodic substrate in the $x$-direction. In the other
direction, periodic boundary condition means that $x_{i,N_{y}+1}$ =
$x_{i,1}$. The average distance between particles, $b$, in each chain
is determined through the boundary condition and is given by $b=N_{s}\
a/N$ and therefore the overlayer coverage, or atomic concentration,
can be defined as $\theta =N/N_{s}=a/b$. With this set up, $\theta $
and $b$ cannot be changed independently. With free boundary conditions
in the $x$-directions, the ratio $b/a$ can assume arbitrary nonrational
values. The system now can remain commensurate for a given $\theta $
only up to a critical value $ \delta _{c}$ of the misfit parameter
\cite{Frenkel} $\delta =(b-a)/a$. This boundary condition is particularly
useful to study incommensurability effects.

An important feature of the model is that the particles have a fixed
number of neighbors. So, even at high temperatures, where a liquid phase
with particles diffusing around independently is usually expected, there is
some crystalline order although the correlations decay as a power law \cite
{Pokrovsky}. We shall, nevertheless, refer to this phase as a ''fluid '',
keeping in mind its distinct nature. At low temperatures, the periodic
potential is able to pin the overlayer and true long-range order prevails
leading to a finite static friction. The low and high temperature phases 
are separated by a commensurate solid to fluid transition at a critical
temperature $T_c$ where the overlayer at high temperatures become disordered 
and is effectively depinned from the substrate. We shall also refer to this
transition as "melting" although topological defects, as dislocations, 
are not allowed due to the fixed-neighbor constraint. In the absence of 
a periodic potential, disorder plays an important role in pinning the 
overlayer \cite{carraro,giamarchi} and glassy behavior is expected.

The periodic potential in Eq.(1) arises from an adiabatic average
over the substrate degrees of freedom. The nonadiabatic excitations
of the substrate provide energy exchange between the substrate and
the adlayer and lead to damping and fluctuation in the motion of the
adatoms. In our model, this effect is  described by the usual Markovian
Langevin dynamics. The substrate acts as a heat bath at an  equilibrium
temperature $T$ and the coupling of the substrate excitations and
adlayer is characterized by a friction parameter $\eta $. Note that
in the absence of  the periodic potential, the adatom interactions do
not affect the collective response of the adlayer and the macroscopic
sliding friction is equal to the microscopic coupling parameter $\eta $.
.

\section{ General Properties of Sliding friction}

To characterize the response of the adlayer to the external driving force, 
we define the sliding friction coefficient $\bar{\eta}$ through the relation

\begin{equation}
v_{d}=F/m\bar{\eta}
\end{equation}
where $v_{d}$ is the average velocity of all the adparticles representing
the drift velocity in the direction of the force.  In general, the
sliding friction $\bar{\eta}$ has a complicated dependence on the
external force $F$ in the nonlinear regime. First, we remark on a number
of qualitative features of the drift velocity $v_{d}$ and the sliding
friction $\bar{\eta}$ that can be readily understood without detailed
calculations \cite{Gallet}. In the presence of the external force F,
the potential is tilted like a washboard with local minima which become
weaker as the strength of the force is increased. Beyond a critical force
value $F_{o}$, the effective potential minima disappears. For the present
model, $F_{o}=2\pi U_{o}$, the same result as for the one-dimensional
case \cite{Pokrovsky}. At $T=0$ and for a commensurate value of coverage
$\theta $ , $v_{d}$ remains zero for any applied force less than $F_{o}$.
When $F>> F_{o}$, the layer slides as a whole with a  sliding friction
$\bar{\eta} \ge \eta$.
At low but finite temperatures, the drift velocity induced by a small
external force is limited by the nucleation and mobility of defects in
the commensurate structure. In one-dimension, these defects are kinks
and antikinks (local compression or extension of the commensurate atomic
configuration) and can always be excited at any finite temperature leading
to a finite $\bar{ \eta}$ at any $T\neq 0$ . In two dimensions, however,
the corresponding excitations are linear defects (domain walls) which
have a nonzero free-energy per unit length for temperatures below the
melting transition temperature $T_{c}$. As thermal fluctuations, these
defects can only be nucleated as closed loops with an activation energy
proportional to its length and therefore do not contribute  significantly
to the linear mobility of the  system, $\lim_{F\rightarrow 0} v_d/F $, in
the thermodynamic limit and one expects that for a pinned solid, 
$1/\bar{\eta}$  is essentially zero for $T<T_{c}$ in this regime. 
When the external force is increased from zero, activated nucleation and 
growth of commensurate domains induced by the external force can contribute  
to a finite mobility, leading to a nonlinear friction 
$\bar{\eta} > \eta$ with a complicated dependence on the external force F. 
Eventually, in the limit of large external force such that the effect of 
the periodic potential is negligible altogether, $\bar{\eta}$
again reduces to the microscopic friction  parameter  $\eta$.

In the nonlinear regime, the behavior of drift velocity $v_{d}$ for
increasing external force $F$ is expected to depend strongly on the
initial phase \cite{Persson} at $F=0$. If the adsorbate, when $F=0$, is a
fluid, $ v_{d}$ will be nonzero for arbitrarily small external force. This
result is also true for the present model. In the fluid phase above
$T_{c}$, the overlayer is essentially depinned and any small force leads
to nonzero $ v_{d} $. Moreover, the relation $v_{d}=f(F)$ should show
no hysteresis. For an initial state at $F=0$ that is incommensurate, the
critical value of the force for finite mobility can be much lower as shown
in previous works of Persson \cite{Persson} and Braun et al.\cite{braun}.

\section{Molecular dynamics simulation}

To analyse the nonlinear sliding friction of the model presented in 
 section II, we have performed a molecular dynamics simulation study
of the Langevin equation resulting from the model. An external  driving
force $F$, representing the effect of the other surface in the boundary
lubrication problem, is assumed to act on each of the adsorbates. The
equation of motion for the particle with coordinate ${x}_{ij}$ is given
by the Langevin equation

\begin{equation}
m\ddot{x}_{ij}+m\eta \dot{x}_{ij}=-{\frac{\partial V}{\partial x_{ij}}}-{ 
\frac{\partial U}{\partial x_{ij}}}+F+f_{ij}  \label{Langevin}
\end{equation}
where $U$ is the periodic potential, $V$ is the adsorbate-adsorbate harmonic
interaction potential and $f_{ij}$ is a random force that is related to the
microscopic friction parameter $\eta $ by the fluctuation dissipation
theorem 
\begin{equation}
<f_{ij}(t)f_{i^{\prime }j^{\prime }}(t^{\prime })>=2\eta mk_{B}T\ \delta
_{i,i^{\prime }}\delta _{j,j^{\prime }}\delta \ (t-t^{\prime })
\end{equation}

We have studied the sliding friction by simulating the above equations
using methods of Brownian molecular dynamics \cite{Allen} which
treat accurately the effect of the systematic and stochastic forces
in the problem. The calculations were performed at a commensurate
coverage $\theta =1/2$ ($b/a=2 $). Typically, the time variable
was discretized with time step $\delta t=0.02-0.06\ \tau $ where
$  \tau =(ma^{2}/U_{o})^{1/2}$ and $4-10\times 10^{5}$ time steps
were used in each calculation allowing $10^{5}$ time steps for
equilibration. Calculations were performed as function of temperature
and external force on systems containing $N\times N$ adsorbate particles,
where typically $N=10$, and the elastic constant was set to $K=10$ .

In the calculations, the system was allowed to evolve into a steady state
such that the time average of physical quantities, like the drift velocity $ 
v_{d}$, approached a constant. The effective temperature $T^{*}$ of the
overlayer during sliding was obtained by two different methods. Using the
fluctuation of particle velocities from the average as

\begin{equation}
kT^{*}=m(<v^{2}>-<v>^{2})
\end{equation}
and by equating the energy transferred to the adsorbed system by the
external force to the energy transferred to the substrate which gives \cite{Persson}

\begin{equation}
\frac{T^{*}}{T}=1+\frac{F^{2}}{m\eta \bar{\eta}T}[1-\frac{\eta }{\bar{\eta}}]
\end{equation}
where $\bar{\eta}=F/mv_{d}$. The two methods agree within the estimated
errors, for calculations expected to be equilibrated in a steady state, and
provide additional support for the assumption that local thermalization of 
the overlayer can be described by an effective temperature $T^*$. 

\section{Results and Discussion}

In this section, we present results from our molecular dynamics
simulations.  We use units in which $a=1$, $m=1$ and $U_{0} =1$ which sets
the scales for length, mass and energy respectively. The corresponding
time scale is $\tau =(ma^{2}/U_{o})^{1/2}$, and the temperature scale
is $ T_{0}=U_{0}/k_{B} $. It is also convenient to normalize the external force
by the zero-temperature critical value $F_{o}$.

To obtain an estimate of the critical temperature $T_c$, we 
calculated $\bar{\eta}$ as a function of temperature for 
a small value of the external force, $F/F_{o}=0.035$. At 
low temperatures, $1/\bar{\eta}$ is essentially zero and 
the adlayer is in a pinned state. As the temperature is 
increased $1 /\bar{\eta}$ remains zero until \cite{granato} 
$T_{c}\sim 2.3$, beyond which it increases significantly and 
should approach $1$ at high enough $T$. The value of $T_{c}$ 
correspond to the commensurate solid-fluid transition. Above $T_c$, the
overlayer is effectively depinned from the substrate. 

\begin{figure}[tbp]
\centering\epsfig{file=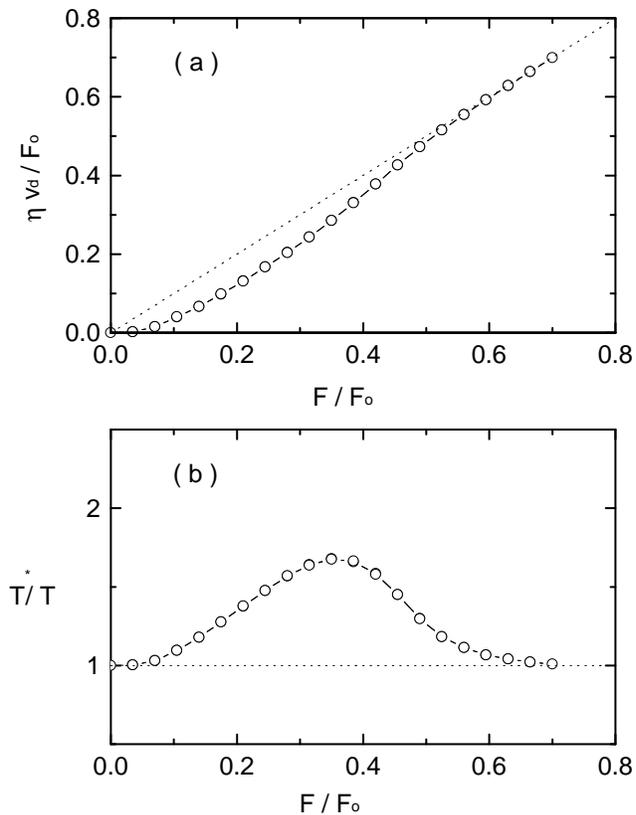,bbllx=6cm,bblly=0.7cm,bburx=16cm,
 bbury=28cm,width=4.5cm}

\caption{Drift velocity $v_d$ and effective temperature $T^*$ 
as a function of external force $F$ for $\eta
= 0.6$ and $T=3$. Dotted lines correspond to free overlayer behavior. }
\label{hystT3}
\end{figure}

\begin{figure}[tbp]
\centering\epsfig{file=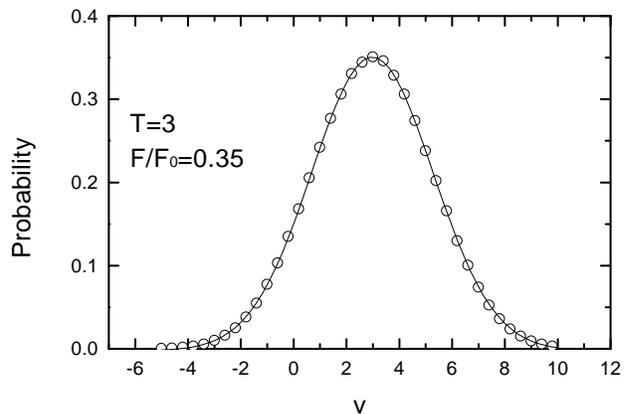,bbllx=6cm,bblly=12cm,bburx=16cm,
 bbury=28cm,width=4.5cm}

\caption{Probability distribution (circles) of particle 
velocities in the stead state for $F/F_o=0.35$ in Fig. \ref{hystT3}. 
Solid line is a Gaussian fit. }
\label{prob}
\end{figure}

In Fig. \ref{hystT3}  we show the behavior of $v_{d}$ 
and the effective temperature of the layer $T^{*}$, obtained
at a temperature $T=3$ which is  above $T_{c}$. In this case, the adlayer is
in a fluid state at zero external force. The results for $F$ increasing
from zero and decreasing from its highest value are the same, showing a
total absence of hysteresis effects. The effective friction coefficient
$\bar{\eta}$ is finite even at vanishingly small values of $F$.
In the absence of the periodic potential, $\bar{\eta}=\eta$ and $v_{d}$ 
should depend linearly on $F$ as indicated by the dotted line 
in Fig. \ref{hystT3}. In the present case, 
the net motion of the overlayer is limited by defect excitations induced by
the coupling to the periodic potential leading to an enhancement 
of $\bar{\eta}$ over the microscopic friction parameter $\eta $ and 
an increase of $T^{*}$ over the equilibrium temperature $T$, as $F$ increases.  
The sliding friction $\bar{\eta}$ approaches $\eta $ only for larger $F$. 
The calculation of the effective temperature $T^*$ assumes that the
particle velocities have a Maxwellian probability distribution of width 
$k_B T^*$. In fact, as shown in Fig. \ref{prob} the velocity distribution 
can be well described by a Gaussian distribution.

\begin{figure}[tbp]
 \centering\epsfig{file=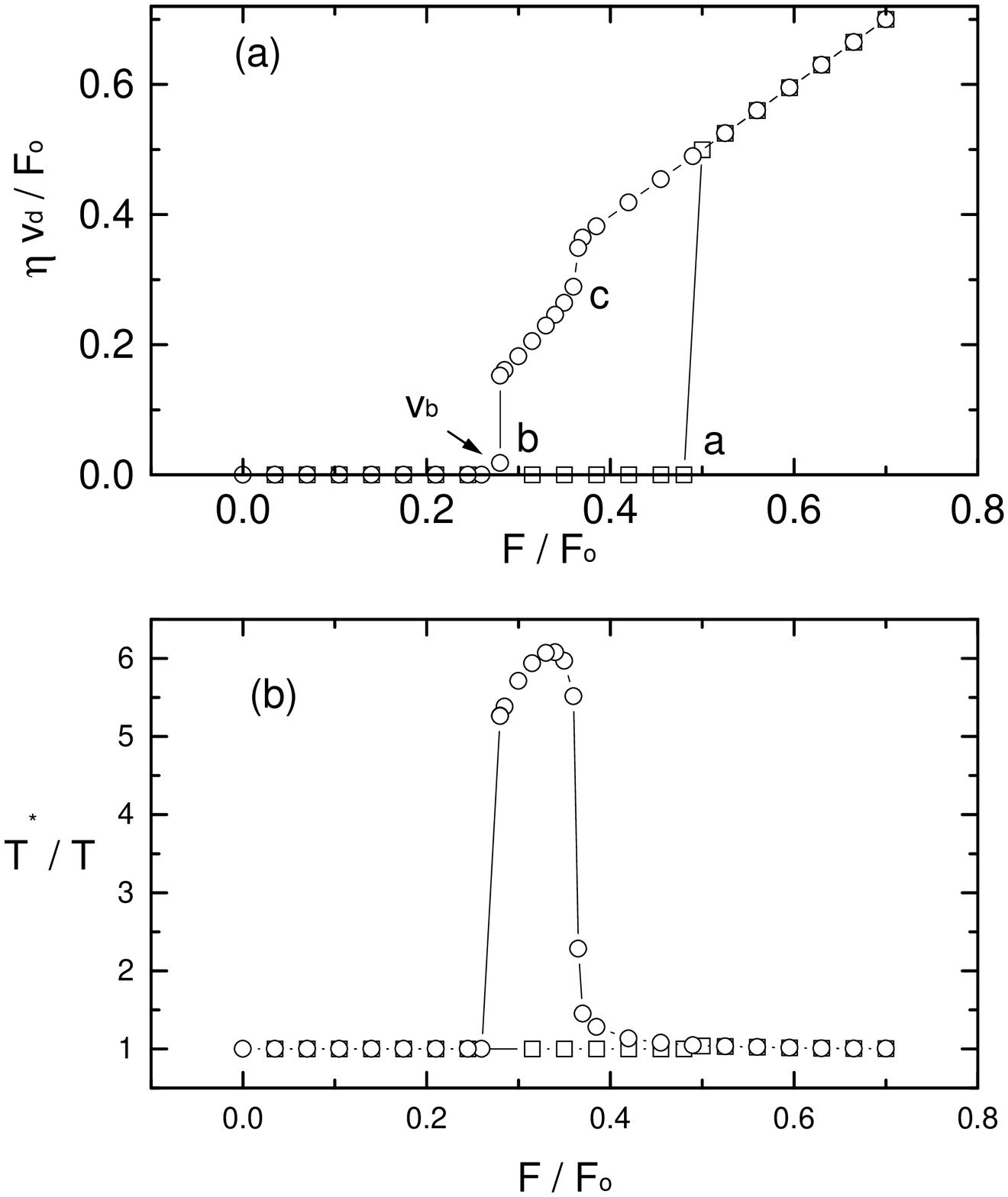,bbllx=6cm,bblly=0.7cm,bburx=16cm,
 bbury=28cm,width=4.5cm}

\caption{Drift velocity $v_d$ and effective temperature $T^*$ as 
a function of external force $F$ for $\eta= 0.6$ and $T=0.5 $ 
and periodic boundary conditions. Squares correspond to data for increasing $F$ and
circles for decreasing $F$. The arrow indicates the minimum velocity $v_b$.}
\label{hystT05}
\end{figure}

For $T<T_{c}$ however, where a commensurate solid phase prevails when
$F=0$, a critical force $F_{a}$ is necessary to depin and initiate
sliding. It was found for the Lennard-Jones system \cite{Persson}
that in this case the nonlinear sliding friction exhibit hysteresis as
a function of $F$, i.e., the relation between velocity and external
force depends on whether $F$ increases from zero or decreases from a
high value. We find the same behavior in our model for sufficiently low
temperatures. In Fig.  \ref{hystT05}, we show the results for $v_{d}$
and the effective temperature of the adlayer $T^{*}$ as a function of $F$
for $ \eta =0.6$ obtained at a temperature $T=0.5$ which is much less
than $T_{c}$. This temperature is also considerably below the amplitude
of the periodic potential $U_{o}$ . As will be discussed below,  we find
that $T<U_{0}$ seems to be necessary in order that a sharp hysteresis
feature appears in the $v_{d}\times F$ characteristics.  The calculation
was performed with the adlayer initially in a pinned state, then the
external force was increased to different values. An hysteresis loop
is clearly seen in Fig. \ref{hystT05} and share the same features as
found in the simulations of the Lennard-Jones systems \cite{Persson}.

\begin{figure}[tbp]
\centering\epsfig{file=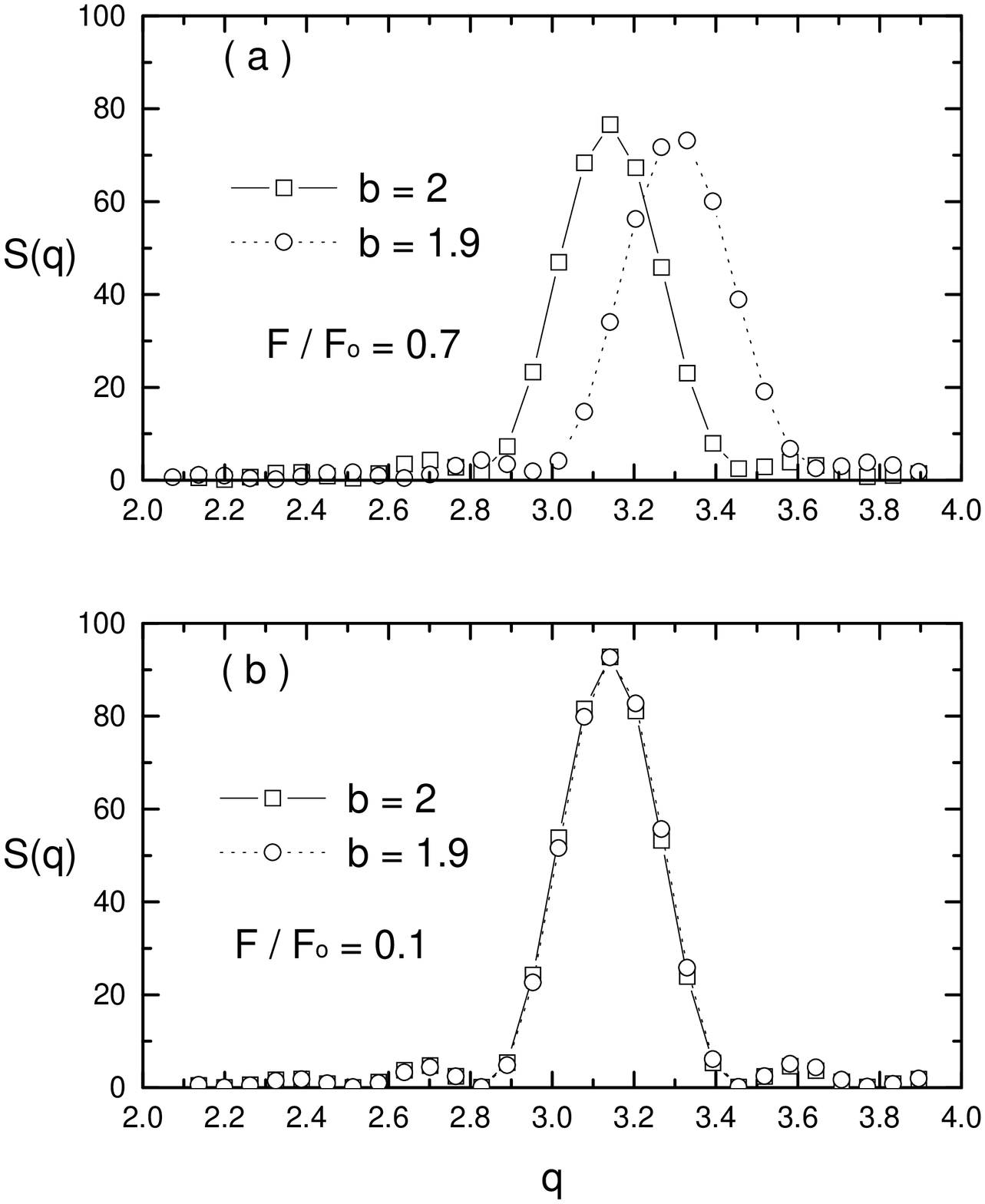,bbllx=6cm,bblly=0.7cm,bburx=16cm,
 bbury=28cm,width=4.5cm}

\caption{Structure factor $S(q)$ for $b/a=2$ and $b/a=1.9$ with $\eta= 0.6$
and $T=0.5 $ and free boundary conditions, in the sliding (a) and pinned (b)
phases. }
\label{sffbT05}
\end{figure}

In Fig. \ref{hystT05}, it is shown that the initial sliding phase, 
when the applied force is increased beyond the static threshold $F_{a}$, 
has almost the same temperature as the substrate and the 
substrate potential provides no additional resistance. 
This is a sliding solid phase. It corresponds to the floating 
or incommensurate solid phase in the equilibrium
situation. It is important to note that this phase is dynamically
generated by the external field since the initial temperature is
below $T_{c}$ and a commensurate solid phase would be the equilibrium
state. With  the choice of periodic boundary conditions in our model, 
the misfit $\delta $ is zero and the configuration of the sliding
incommensurate solid phase in Fig. \ref{hystT05} is indistinguishable
from a commensurate phase. However, using free boundary conditions and
varying the misfit parameter we can confirm that this phase corresponds to
a sliding solid phase incommensurate with the substrate. 
In Fig. \ref {sffbT05} we show the structure factor $S(q)$ for 
two values of the overlayer lattice spacing, $b/a=2$ and $b/a=1.9$ , 
and at two different limits of the external force, $F/F_{o}=0.1$ 
in the pinned phase (Fig. \ref {sffbT05}b) and $F/F_{o}=0.7$ in 
the sliding phase (Fig. \ref{sffbT05}a). In the sliding 
phase, $S(q)$ is peaked at the wavevector $2\pi/b$ independent of
the substrate while it remains locked to the commensurate value $\pi $ in
the pinned phase. Thus the external force induces a pinned commensurate
solid to sliding incommensurate solid transition at $F=F_{a}$. As the
force is decreased from its highest value through $F_{a}$, there is a
jump in $T^{*}$ accompanied by a simultaneous decrease in $v_{d}$ at a
critical force $F_{c}$. This corresponds to an overheating of overlayer
to an effective temperature which is above $T_{c}$,
dynamically ``melting'' the sliding solid phase. This can be regarded as a
sliding solid to fluid transition. Finally at $F=F_{b}$, the fluid phase
condenses into a pinned commensurate solid phase and the temperature of
the adlayer and the substrate become the same again. The fact that there
is a condensation from the ''melted'' phase is confirmed by noting that
the temperature of the adlayer at this condensation is almost identical
to the equilibrium critical temperature $T_{c}$. Persson \cite{Persson}
has observed very similar dynamical transitions for Lennard-Jones systems,
showing that these features are universal in the non-linear response of
overlayers and not just for specific models. In fact, other versions of
the Frenkel-Kontorova model also show some of these features \cite{braun}. 
In addition the present results show that a conventional ''liquid''
phase is not required and a high temperature thermally disordered phase
with nonzero linear mobility is sufficient to show some of these effects.

\begin{figure}[tbp]
\centering\epsfig{file=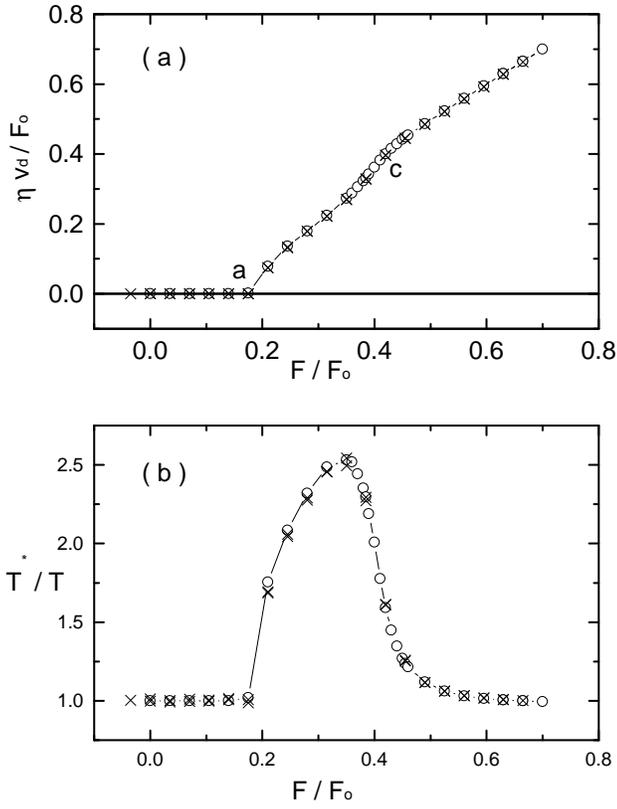,bbllx=6cm,bblly=0.7cm,bburx=16cm,
 bbury=28cm,width=4.5cm}
\caption{Drift velocity $v_d$ and effective temperature $T^*$ 
as a function of external force $F$ for $\eta
= 0.6$ and $T=1.5$. Circles and crosses correspond to 
calculations using equilibration times differing by a factor of 10. }
\label{hystT15}
\end{figure}

\begin{figure}[tbp]
 \centering\epsfig{file=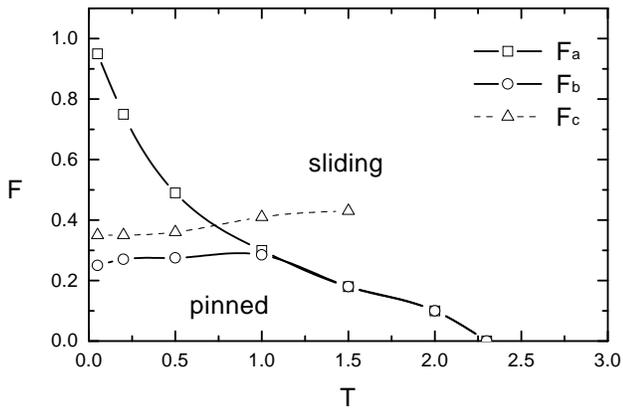,bbllx=6cm,bblly=12cm,bburx=16cm,
 bbury=27cm,width=4.5cm}

\caption{Critical forces $F_a$, $F_b$ and $F_c$ as a function of temperature
for $\eta=0.6$. }
\label{phd}
\end{figure}

We now consider the behavior of the nonlinear sliding friction at a
higher temperature $T=1.5$ but still less than $T_{c}$. As can be
seen from Fig.  \ref{hystT15}, the hysteresis disappears but $F_{a}$
can still be defined.  In addition, a smooth sliding fluid to solid
transition takes place at $ F_{c}>F_{a}$. At this higher temperature,
the system is able to relax in a short enough time to a stable single
velocity, independent of the history of force variation. The behavior 
does not depend significantly on the rate of change of the force 
since calculations with different equilibration times give the same 
result as indicated in the figure. The hysteresis observed at low 
temperature is also not affected significantly.  However, we can not 
exclude that for large enough equilibration times the range of 
temperature where hysteresis is observed is much smaller.
The behavior of $F_{a}$ , $F_{b}$ and $F_{c}$ as a function of 
temperature is shown in Fig. \ref{phd}. Within the present model, 
hysteresis disappears at $T\sim U_{0}<T_{c}$ in contrast with 
the arguments  \cite {Persson} suggesting that smooth 
sliding only occurs  above $T_{c}$.

\begin{figure}[tbp] 
 \centering\epsfig{file=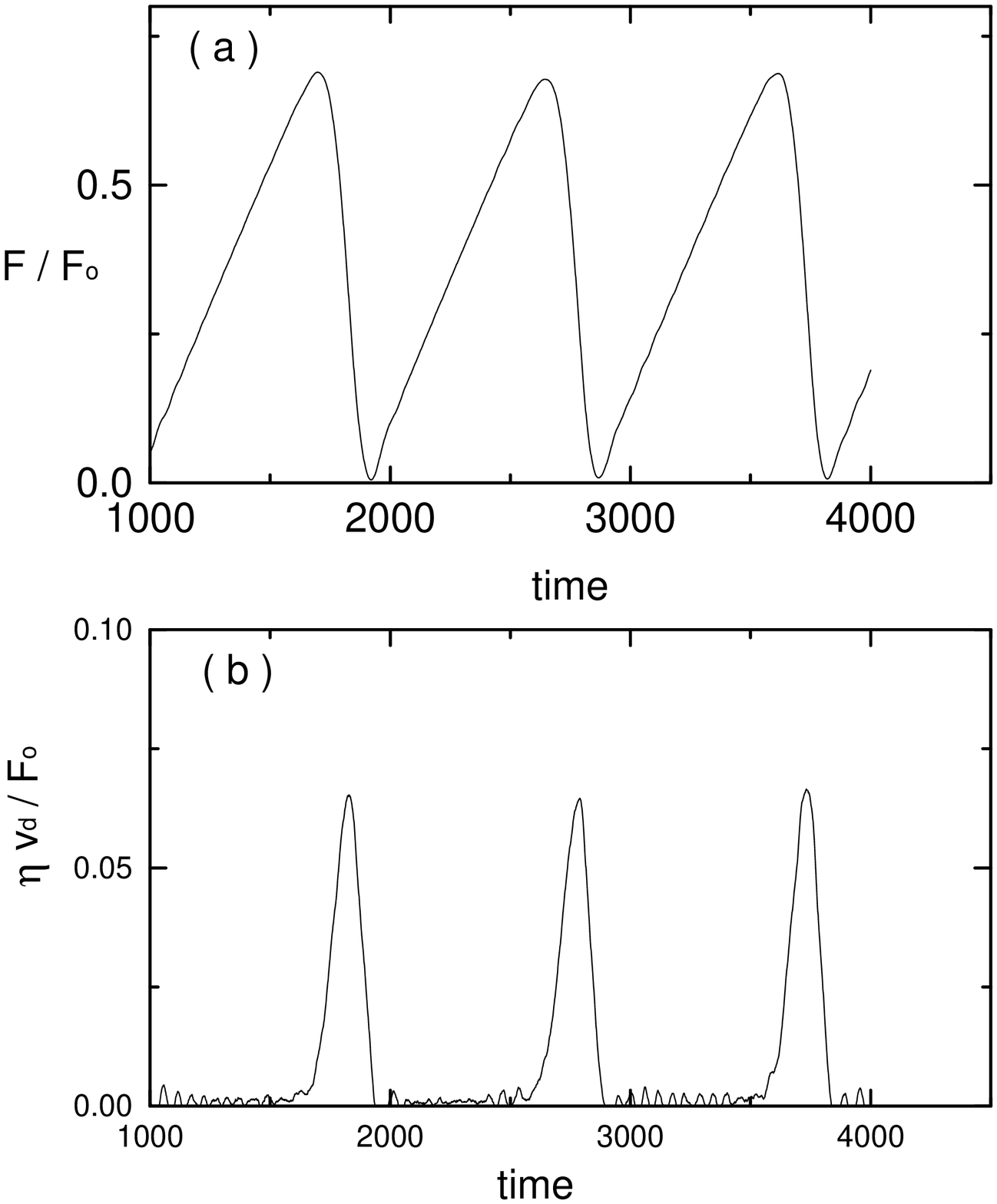,bbllx=6cm,bblly=0.7cm,bburx=16cm,
 bbury=28cm,width=4.5cm}

\caption{Time evolution of the $v_d$ and $F$ when the system is pulled by a
spring at constant velocity $v_s=0.01 $, below $v_b$, for $\eta = 0.6$, $ 
T=0.5$ and $K_s = 0.5$.}
\label{stick}
\end{figure}

\begin{figure}[tbp]
 \centering\epsfig{file=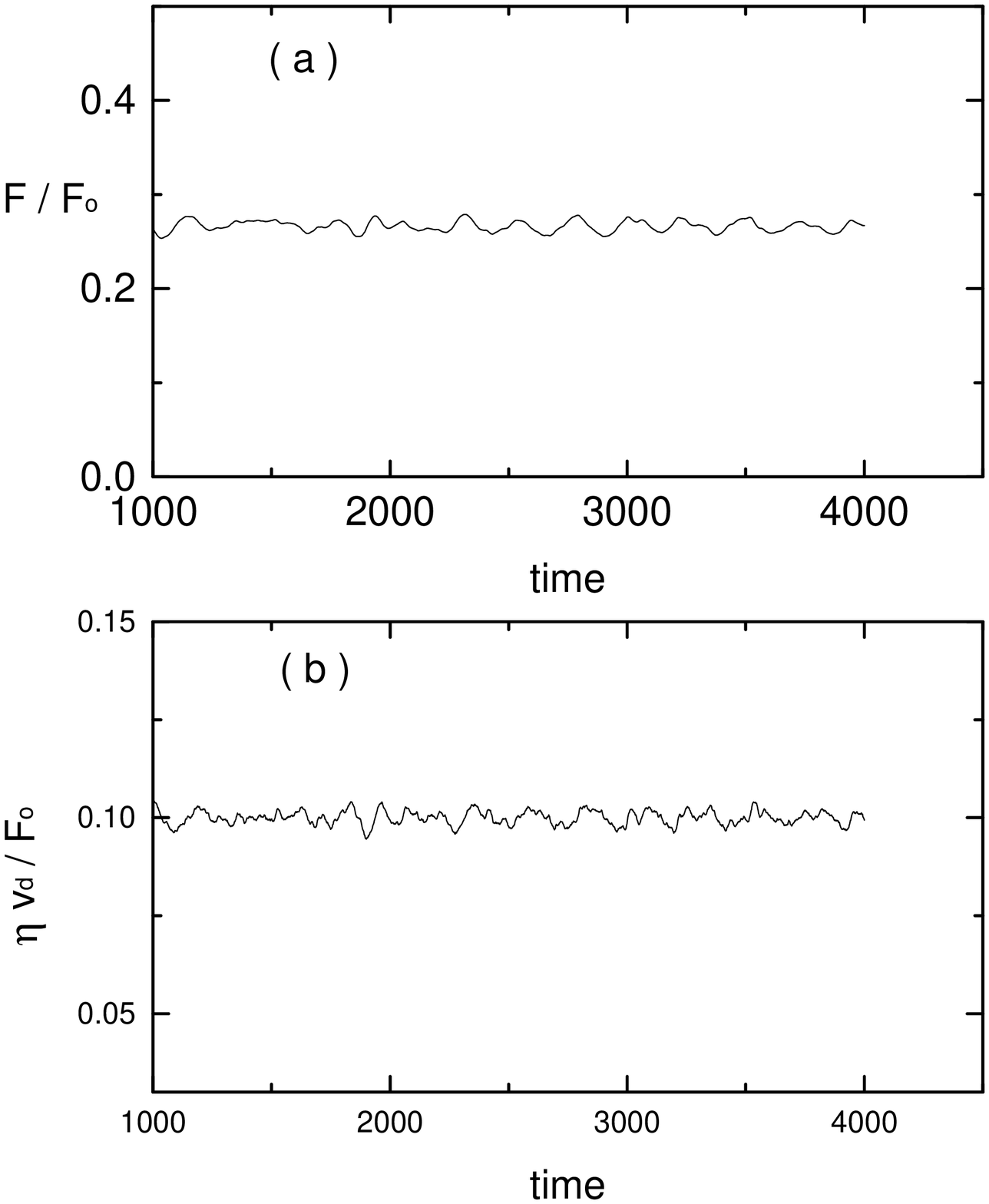,bbllx=6cm,bblly=0.7cm,bburx=16cm,
 bbury=28cm,width=4.5cm}

\caption{Time evolution of the $v_d$ and $F$ when the system is pulled by a
spring at constant velocity $v_s=0.1$, above $v_b$, for $\eta = 0.6$, $T=0.5$
and $K_s=0.5$.}
\label{smooth}
\end{figure}

An important feature of the $F\times v_{d}$ curve in Fig. \ref{hystT05}
is the existence of a ''velocity gap '', $0<v<v_{b}$ , which leads to
stick-and-slip motion \cite{Persson}. In the gap region, the overlayer
cannot sustain a  constant  sliding velocity and stick-and-slip motion is
expected when it is driven by a spring  moving with constant velocity
lying in the gap, as  measured in sliding friction experiments
\cite{Israelachvili}. Smooth sliding is  possible only for $v_{s}>v_{b}$. 
We have confirmed this behavior by calculating the time dependent drift
velocity $v_{d}(t)$ and force $F(t)$ when each particle of the overlayer
is subject to an elastic external force of the form

\begin{equation}
F(t)=K_{s}(x_{cm}-v_{s}t)
\end{equation}
where $K_{s}$ is the spring elastic constant and $x_{cm}$ is the position of
the center of mass of the overlayer. Indeed,  stick and
slip motion is obtained for $v_{s}<v_{b}$ as shown in Fig. \ref{stick} and 
smooth sliding motion  for $v_{s}>v_{b}$ as shown in Fig. \ref{smooth}. 

\begin{figure}[tbp]
 \centering\epsfig{file=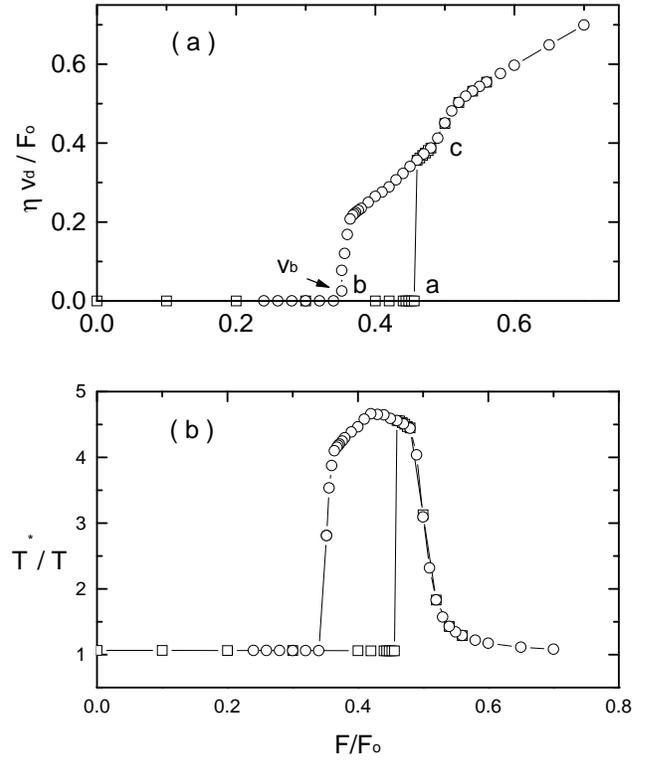,bbllx=6cm,bblly=0.7cm,bburx=16cm,
 bbury=28cm,width=4.5cm}

\caption{Drift velocity $v_d$ and effective 
temperature $T^*$ as a function of external force $F$ for $\eta
= 0.9$ and $T=0.5 $. Squares correspond to data for increasing $F$ and
circles for decreasing $F$. The arrow indicates the minimum velocity $v_b$.}
\label{hyste09}
\end{figure}

\begin{figure}[tbp]
\centering\epsfig{file=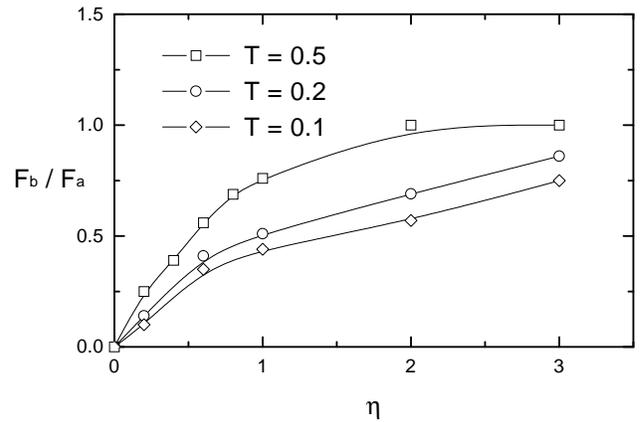,bbllx=6cm,bblly=12cm,bburx=16cm,
 bbury=27cm,width=4.5cm}

\caption{Ratio $F_b/F_a$ as a function of $\eta $ and $T$. }
\label{ratio}
\end{figure}

We have also studied the effects of varying the microscopic friction
parameter $\eta $ on the nonlinear mobility. The $F\times v_{d}$
characteristics  is shown in Fig. \ref{hyste09} for a larger value of
$\eta =0.9$ . In this case, the sliding fluid to solid transition
occur at $F_{c}>F_{a}$ and the return to the pinned solid phase at
$F_{b}$ occurs at an effective temperature of the overlayer $T^{*}$
below $T_{c}$, so the results are qualitatively different from that for
$\eta =0.6$. Persson \cite{Persson} has argued, based on hydrodynamical
considerations that whenever $T^{*}<T_{c}$ is satisfied at $F_{b}$ ,
the ratio $F_{b}/F_{a}$ between the kinetic and static friction forces
should be a universal value $\sim 1/2$. For the present model, we find
instead that the ratio $F_{b}/F_{a}$ depends on $\eta $ and $T$ as shown
in Fig.\ \ref{ratio}. The ratio approaches $1$ for $ \eta \sim3$  at
$T=0.5$ . At lower temperatures, the results indicate that the ratio
would asymptotically  approach one at larger values of  $\eta $. This
behavior is expected since in the large $\eta $ limit the system is
overdamped and hysteresis is not expected. In fact, we have confirmed
through direct calculation that when the inertial term is dropped in  the
equation of motion Eq. \ref {Langevin},  hysteresis effects disappear and
$F_{b}/F_{a}=1$ even for the lowest temperature in Fig. \ref{ratio}. This
should be contrasted to the results obtained by Persson \cite{Persson}
for the Lennard-Jones system where $ F_{b}/F_{a}$ seems to saturate for
large $\eta $ at $\approx 0.6$ for temperatures below $T_{c}$ .

\begin{figure}[tbp]
 \centering\epsfig{file=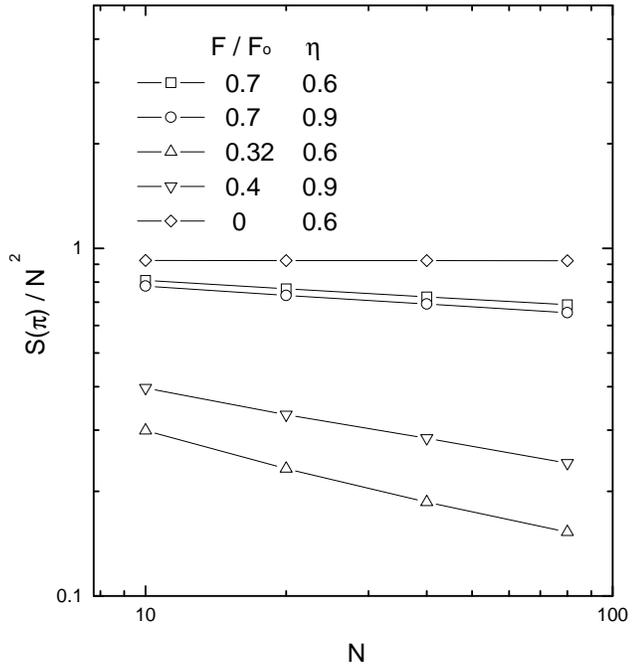,bbllx=6cm,bblly=0.7cm,bburx=16cm,
 bbury=26cm,width=4.5cm}

\caption{Finite-size behavior of the peak in the structure factor $S(q)$ for
different values of $F$ in Figs. \ref{hystT05} and \ref{hyste09}. }
\label{psf}
\end{figure}

We should note however that the nature of the high-temperature phase in the FK
model is quite different from the Lennard-Jones system, with power law
instead of the exponential decay of density correlations as a function
of separation distance $r$. This follows from the fixed-neighbor constraint
in the model of Eq. \ref{hamilt} which excludes topological defects, 
such as dislocations. It is well known that thermally excited dislocations can 
destroy crystalline order and lead to exponential decay of density 
correlations \cite{Pokrovsky}. Therefore, in the absence of dislocations, the equilibrium
high temperature phase for $T > T_c$ and $F=0$ is a depinned elastic 
solid and correlations decay as a power law. In the sliding state 
for $F>0$, one thus expect  that when the effective temperature 
of the overlayer $T^*$ is larger 
than  $T$, overheating of the overlayer leads to power-law decay 
of density correlations,  whereas long-range order should occur 
when $T^*=T$.  This expected  behavior is confirmed in Fig. \ref{psf} 
where the finite-size behavior of the structure factor peak for 
different values of $F$ and $\eta $ is shown, corresponding to 
the different dynamical phases in Figs. \ref{hystT05} 
and \ref{hyste09}. For density correlations that decay
algebraically as $r^{-\eta _{o}}$, the normalized peak height of $S(q)$
at the commensurate wavevector $q=\pi $ should scale as a power law $S(\pi
)/N^{2}\propto N^{-q^{2}\eta _{o}}$  while it should be a constant
when long-range order prevails \cite{Pokrovsky}. For the fluid phases where
$T^* >T$, corresponding to the value of $F/F_{o}=0.32$ in Fig. \ref {hystT05}
and $F/F_{o}=0.4$ in Fig. \ref{hyste09}, the loglog plot $S(\pi )$
vs $N$ has an apparent linear behavior with nonzero slope as shown in
Fig. \ref{psf}, consistent with algebraic order. For the pinned solid
phase where $T^*=T$ at $F=0$, the  quantity $S(\pi ))/N^{2}$ 
is a constant indicating long range order. For the sliding 
solid phase at $F/F_{o}=0.7$, the
corresponding value for $S(\pi ))/N^{2}$ decreases very slowly with
increasing $N$, indicating that it is  not a phase with full long range
order in contrast with the pinned solid phase at $F=0$ but 
much larger system sizes would be required to fully confirm this behavior.  

The hydrodynamics arguments
used by Persson \cite{Persson} suggesting a universal value $F_{b}/F_{a}$
$\sim 0.5$ are based on the existence of a drag force acting on a nucleating
domain of solid phase surrounded by the fluid phase. 
Since the nature of the fluid phases in the FK
and Lennard-Jones models are very different, this argument is not  valid 
for the FK model and this could account for the different behavior of 
the ratio $F_{b}/F_{a}$. However, in view of the
many common features shared by  the two models as shown in Fig. \ref{hystT05}
and \ref{hyste09}, it is also possible that the relation $F_{b}/F_{a}$
$\sim 0.5$ is in fact valid only within a limited range of $\eta $ and
$T$, as shown in Fig.\ref{ratio}, and the Lennard-Jones system simply has a
broader crossover region than the present model.

\section{Conclusions}

Comparing our results for the nonlinear response for the Frenkel-Kontorova
model to the earlier study based on the Lennard-Jones interaction
\cite{Persson}, we found that all the important features such as the
appearance of dynamic sliding solid and fluid phases, the transition
between these phases , the strong hysteresis effects, and the velocity
gap for constant velocity sliding  are common to both models. This
indicates that these are robust features that does not depend on the
details of the model such as  many body interactions among the adatoms
and surface geometry or anisotropies. The strong qualitative similarity
of the results for mobility in these models to the effects observed
in macroscopic sliding between two surfaces support strongly the idea
that the response of the lubricant layer to an external force is the
key factor in controlling boundary lubrication.

Besides the relevance to the  macroscopic sliding friction, the dynamic
phases generated by the external force are of intrinsic interest and
have no counterparts in equilibrium. They are certainly worthy of more
detailed study both experimentally and theoretically and may also be
relevant to other types of driven lattice systems \cite{giamarchi}
as for vortex lattices in type II superconductors.

\section{Acknowledgment}

This work was supported in part by a joint NSF-CNPq grant, an ONR grant
(S.C.Y.) and by FAPESP (Proc. 97/07250-8) (E.G). We thank M.R. Baldan 
for helpful discussions.

\newpage

\end{document}